\documentclass[AMA,Times1COL]{WileyNJDv5} 

\articletype{Research Article}%

\received{Date Month Year}
\revised{Date Month Year}
\accepted{Date Month Year}
\journal{Preprint}

\usepackage{color}
\usepackage{booktabs}
\usepackage{amsmath}
\usepackage{multirow}
\usepackage{tabularx}
\usepackage{graphicx}
\usepackage{caption}
\makeatletter
\long\def\@makecaption#1#2{%
 \vskip3pt
 \sbox\@tempboxa{\normalsize{\bf #1}\quad #2}%
 \ifdim \wd\@tempboxa >\hsize
 \normalsize #1\quad #2\par
 \else
 \global \@minipagefalse
 \hb@xt@\hsize{\box\@tempboxa}%
 \fi
 \vskip4pt
 }
\makeatother
\graphicspath{{figs/}}

\begin{document}

\title{Wavelet Packet-Based Diffusion Model for Ground Motion Generation with Multi-Conditional Energy and Spectral Matching}

\author[1]{Yi Ding}
\author[2]{Su Chen}
\author[1,3]{Jinjun Hu}
\author[2]{Xiaohu Hu}
\author[2]{Qingxu Zhao}
\author[2,4]{Xiaojun Li}

\authormark{Yi Ding et al.}
\titlemark{Wavelet Packet-Based Diffusion Model for Ground Motion Generation with Multi-Conditional Energy and Spectral Matching}

\address[1]{\orgdiv{State Key Laboratory of Precision Blasting}, \orgname{Jianghan University}, \orgaddress{\state{Wuhan}, \postcode{430056}, \country{China}}}

\address[2]{\orgdiv{Key Laboratory of Urban Security and Disaster Engineering of the Ministry of Education}, \orgname{Beijing University of Technology}, \orgaddress{\state{Beijing}, \postcode{100124}, \country{China}}}

\address[3]{\orgdiv{Key Laboratory of Earthquake Engineering and Engineering Vibration}, \orgname{China Earthquake Administration}, \orgaddress{\state{Harbin}, \postcode{150080}, \country{China}}}

\address[4]{\orgdiv{Institute of Geophysics}, \orgname{China Earthquake Administration}, \orgaddress{\state{Beijing}, \postcode{100081}, \country{China}}}

\corres{Su Chen (Email: chensuchina@126.com)\\
Jinjun Hu (Email: hu-jinjun@163.com)}


\abstract[Abstract]{Temporal energy distribution strongly affects nonlinear structural response and cumulative damage. We propose a multi-condition diffusion framework for ground-motion synthesis that simultaneously matches temporal-energy evolution and target response spectra. Wavelet packet decomposition is used to represent the ground-motion signals and enables direct waveform reconstruction via orthogonal filter banks. A Transformer-based conditional encoder with cross-attention integrates heterogeneous conditions, including spectral ordinates, Arias intensity, temporal parameters, and Husid curves. The framework adopts the Elucidating Diffusion Model (EDM) with second-order Heun sampling to improve inference efficiency without sacrificing quality. Tests on the NGA-West2 database show that explicit temporal-energy constraints markedly improve control of energy onset and significant duration while preserving response-spectrum agreement across repeated stochastic samples. The framework yields spectrum-compatible ground motions with realistic energy evolution and enables conditional sample variability to be quantified through repeated stochastic sampling.  A preprint version of this manuscript is available on arXiv under the same title (arXiv:2602.07405; \href{https://doi.org/10.48550/arXiv.2602.07405}{https://doi.org/10.48550/arXiv.2602.07405})}


\maketitle

\section*{Highlights}
\begin{itemize}
\item Integrates response spectra, temporal-energy parameters, and Husid curves for ground-motion synthesis
\item Enables direct waveform reconstruction from wavelet packet coefficients via orthogonal filter banks
\item Uses a variance-exploding EDM with a second-order Heun sampler for efficient, accurate generation
\end{itemize}

\section{Introduction}

Performance-based seismic design and assessment require large sets of ground-motion records for nonlinear time-history analysis to evaluate structural dynamic responses and damage potential under various seismic hazard scenarios \citep{Baker_2006_SpectralShapeEpsilon, Lin_2013_ConditionalSpectrumbasedGround}. Although existing strong-motion databases contain many observed records \citep{Ancheta_2014_NGAWest2Database}, the available records remain unevenly distributed across parameter space. For engineering scenarios requiring specific combinations, such as large-magnitude near-field conditions, particular site classifications under maximum considered earthquakes, or simultaneous satisfaction of multiple intensity-measure constraints, suitable recorded motions are often unavailable.

Early spectral matching methods employed frequency-domain approaches. These methods adjust the Fourier amplitude spectrum based on the ratio of the target response spectrum to the time-history response spectrum, while keeping the Fourier phase of the reference time-history unchanged \citep{Scanlan_1974_EarthquakeTimeHistories}. Another approach adds wavelets in the time domain to the initial time history, which often preserves the nonstationary characteristics of the reference time history \citep{Kaul_1978_SpectrumconsistentTimehistoryGeneration,AlAtik_2010_ImprovedMethodNonstationary, Zhang_2025_GroundMotionConstruction}. With advances in nonstationary signal analysis, methods such as wavelet transform \citep{Spanos_2004_EvolutionarySpectraEstimation, Wang_2022_SimulationFullyNonstationary, Genovese_2025_WaveletbasedGenerationFully}, wavelet packet transform \citep{Yamamoto_2013_StochasticModelEarthquake, Huang_2022_WaveletbasedStochasticModel, Wang_2024_StochasticSimulationPulselike}, and Hilbert-Huang transform \citep{Ni_2011_ApplicationHilberthuangTransform, Ni_2011_TridirectionalSpectrumcompatibleEarthquake} have been introduced to better describe the nonstationarity of ground-motion amplitude and frequency. Generating ground motions that are compatible with a target response spectrum is an inherently ill-posed inverse problem. Mathematically, multiple distinct acceleration time histories may satisfy the same spectral target while differing significantly in physical characteristics such as amplitude envelope, duration, and temporal-energy evolution. Therefore, ground-motion selection and synthesis should consider multiple intensity measures \citep{Huang_2017_EnergycompatibleSpectrumcompatibleECSC, Jin_2023_NewGroundmotionSimulation}.

Recent advances in deep learning have opened new pathways for ground-motion generation. Generative models learn probabilistic mappings from control parameters to time-series samples by modeling high-dimensional stochastic distributions \citep{Kingma_2013_AutoencodingVariationalBayes, Goodfellow_2014_GenerativeAdversarialNetworks, Ho_2020_DenoisingDiffusionProbabilistic}, thereby avoiding the iterative fitting procedures used in traditional methods. Physics-based numerical simulation of ground motions is still constrained to low-frequency bands because of computational cost and limited characterization of small-scale medium heterogeneity. Researchers have therefore explored neural networks that generate broadband motions from low-frequency simulations \citep{Paolucci_2018_BroadbandGroundMotions, Gatti_2020_BlendingPhysicsbasedNumerical, Aquib_2024_BroadbandGroundmotionSimulations}. However, these approaches still depend on random phase generation or the accuracy of low-frequency simulations, limiting control of high-frequency waveform characteristics. Generative adversarial networks (GANs) and their variants have been used for end-to-end waveform synthesis conditioned on source, path, and site parameters \citep{Florez_2022_DatadrivenSynthesisBroadband, Esfahani_2023_TFCGANNonstationaryGroundmotion, Huang_2024_GroundmotionSimulationsUsing}. Related studies have generated response-spectrum-compatible acceleration records \citep{Kim_2024_GenerativeAdversarialNetwork, Miao_2024_ResponsecompatibleGroundMotion, Hu_2026_SpectrumcompatibleArtificialAccelerograms} and synthesized aftershock time histories from mainshock records \citep{Shen_2024_AftershockGroundMotion, Xu_2024_MainshockAftershockSequence}. Variational autoencoders (VAEs) have also been introduced to this domain. Ren et al. \citep{Ren_2024_LearningPhysicsUnveiling} learned amplitude information in the time-frequency domain while implicitly modeling wave propagation through geospatial coordinates and reported more stable training than with GANs.

As an emerging generative framework, diffusion models have produced high-fidelity and diverse samples in computer vision and audio processing \citep{Ho_2020_DenoisingDiffusionProbabilistic, Song_2022_DenoisingDiffusionImplicit, Rombach_2021_HighresolutionImageSynthesis, Zhang_2023_AddingConditionalControl}. Recent research has used diffusion models to learn complex dynamical systems, enabling reconstruction and prediction of full spatiotemporal dynamic fields \citep{Du_2024_ConditionalNeuralField, Li_2024_LearningSpatiotemporalDynamics, Gao_2024_BayesianConditionalDiffusion, Wang_2025_GenerativeSubsurfaceFlow}. Compared with GANs, diffusion models can be trained more stably, accommodate multimodal conditioning, and generate samples that balance diversity and fidelity. With explicit noise-prediction objectives and progressive denoising, diffusion models capture complex ground-motion distributions and produce samples whose statistics closely match observations. Recent studies have applied diffusion models to ground-motion generation, including denoising diffusion probabilistic models (DDPMs), multilabel conditional embedding, and latent-space diffusion, to synthesize high-resolution seismic waveforms \citep{Bergmeister_2024_HighResolutionSeismic, Bi_2025_AdvancingDatadrivenBroadbanda, Huang_2025_GroundmotionGenerationsUsing, Jung_2025_BroadbandGroundMotion}.
Nevertheless, most approaches rely on scalar conditions such as magnitude and distance, providing limited control over temporal-energy distribution and other nonstationary characteristics. For nonlinear structural response analysis, ground motions sharing an identical response spectrum but differing in energy-evolution processes can induce markedly different damage accumulation effects.

To address these limitations, this study proposes a multi-condition ground-motion generation framework based on the Elucidating Diffusion Model (EDM) of Karras et al. \citep{Karras_2022_ElucidatingDesignSpace}. This framework jointly constrains spectral and temporal-energy characteristics using a condition set comprising the response spectrum, Arias intensity, temporal-energy parameters, and the Husid curve. Wavelet packet decomposition maps waveforms into multiscale time-frequency representations. Based on orthogonal filter banks, time-domain waveforms are reconstructed directly from generated wavelet coefficients, avoiding iterative phase retrieval. A lightweight Transformer-based conditional encoder \citep{Vaswani_2017_AttentionAllYou} models dependencies among heterogeneous conditioning variables, and the encoded information is injected into the denoising process through cross-attention in the U-Net architecture \citep{Ronneberger_2015_UnetConvolutionalNetworks}. During inference, a second-order Heun sampler improves efficiency while maintaining generation quality. We validate the method by evaluating intensity-measure control accuracy, time-domain waveform similarity, and conditional sample variability.

\section{Methodology}

\begin{figure*}
 \centering
 \includegraphics[width=0.9\textwidth]{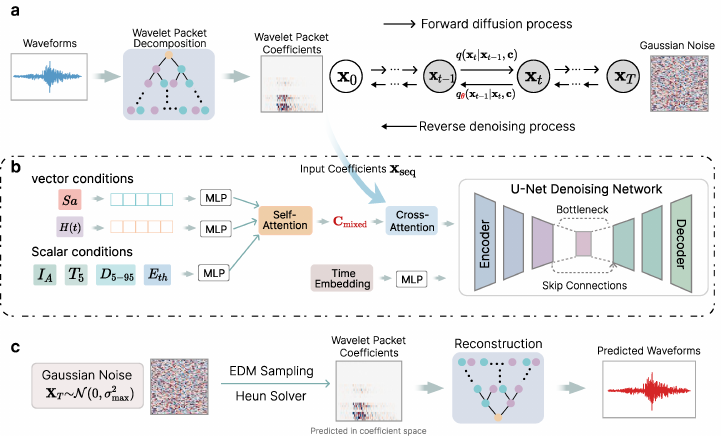}
 \caption{Multi-condition diffusion framework for constrained ground-motion generation: (a) wavelet packet
 decomposition and diffusion in coefficient space; (b) conditional encoder fuses vector conditions ($S_a$,
 $H(t)$) and scalar IMs ($I_A$, $T_5$, $D_{5-95}$, $E_{th}$), and injects them into a U-Net denoiser via
 cross-attention; (c) EDM sampling (Heun solver) from Gaussian noise followed by inverse wavelet packet
 reconstruction to obtain time-domain waveforms.}
 \label{fig:framework}
\end{figure*}

Figure~\ref{fig:framework} summarizes the multi-condition diffusion framework for generating constrained seismic waveforms. The framework comprises three components:
\begin{itemize}
\item Wavelet packet decomposition before training and inverse reconstruction after sampling.
\item A Transformer-based encoder that models dependencies among heterogeneous conditioning variables, with cross-attention in multiscale attention layers.
\item During training, the U-Net denoiser takes the noise-perturbed wavelet packet coefficient matrix, the noise level $\sigma$, and the encoded condition tokens as inputs and predicts the clean coefficient matrix. During inference, the second-order Heun sampler starts from Gaussian noise in coefficient space, and the final coefficient matrix is converted directly into an acceleration time history through the inverse wavelet packet transform.
\end{itemize}

\subsection{Wavelet Packet Representation}
\label{subsec:wavelet}

Generative models must capture long-range temporal dependencies and transient local details in complex waveforms. Training directly on long, high-dynamic-range seismic time series can impair numerical stability and generalization. To capture nonstationary characteristics and operate in continuous representation spaces, previous studies \citep{Esfahani_2023_TFCGANNonstationaryGroundmotion, Bergmeister_2024_HighResolutionSeismic, Xu_2024_HighresolutionGroundMotion} have typically used short-time Fourier transform (STFT) spectrograms as learning targets and reconstructed waveforms using phase-retrieval methods such as the Griffin-Lim algorithm.

 This work uses the Daubechies-6 (Db6) wavelet for a multiresolution wavelet packet transform (WPT) \citep{Daubechies_1988_OrthonormalBasesCompactly, Coifman_1992_EntropybasedAlgorithmsBest}. Its orthogonality, compact support, and six vanishing moments provide a practical balance between temporal localization and frequency resolution for transient ground motions. The corresponding orthogonal filter bank permits direct reconstruction of time-domain waveforms from generated wavelet coefficients without iterative phase recovery. For an acceleration sequence $a[n]$ containing 16,384 samples, a 7-level Db6 decomposition yields $2^7 = 128$ frequency subband nodes with 128 coefficients per node. The resulting $128 \times 128$ representation matches the input geometry of the two-dimensional U-Net. The recursive decomposition is implemented through low-pass and high-pass filters:
\begin{equation}
\text{WP}_{\ell+1,2k}[m] = \sum_n h[2m - n] \text{WP}_{\ell,k}[n], \quad \text{WP}_{\ell+1,2k+1}[m] = \sum_n g[2m - n] \text{WP}_{\ell,k}[n],
\label{eq:wavelet_decomp}
\end{equation}
where $h[n]$ and $g[n]$ denote the low-pass and high-pass decomposition filters of the Db6 wavelet, respectively, $m$ represents the time index after downsampling, $\ell$ denotes the decomposition level, and $k$ indexes nodes at each level.

We normalize coefficients at each node independently. For the $k$-th node coefficients $\mathbf{w}_k$, normalization yields $\tilde{\mathbf{w}}_k = (\mathbf{w}_k - \mu_k)/(\sigma_k + \epsilon)$, where $\mu_k$ and $\sigma_k$ denote the mean and standard deviation, respectively, and $\epsilon = 10^{-6}$ prevents division by zero.

During inference, waveforms are reconstructed using the inverse wavelet packet transform. Starting from the 128 nodes at level 7, synthesis proceeds layer by layer through the reconstruction filters $\tilde{h}[n]$ and $\tilde{g}[n]$:
\begin{equation}
\text{WP}_{\ell-1,k}[n] = \sum_m \left( \tilde{h}[n - 2m] \text{WP}_{\ell,2k}[m] + \tilde{g}[n - 2m] \text{WP}_{\ell,2k+1}[m] \right).
\label{eq:wavelet_recon}
\end{equation}
Here, the synthesis filters $\tilde{h}[n]$ and $\tilde{g}[n]$ satisfy the perfect-reconstruction conditions with the corresponding decomposition filters.

\begin{figure}
 \centering
 \includegraphics[width=0.9\textwidth]{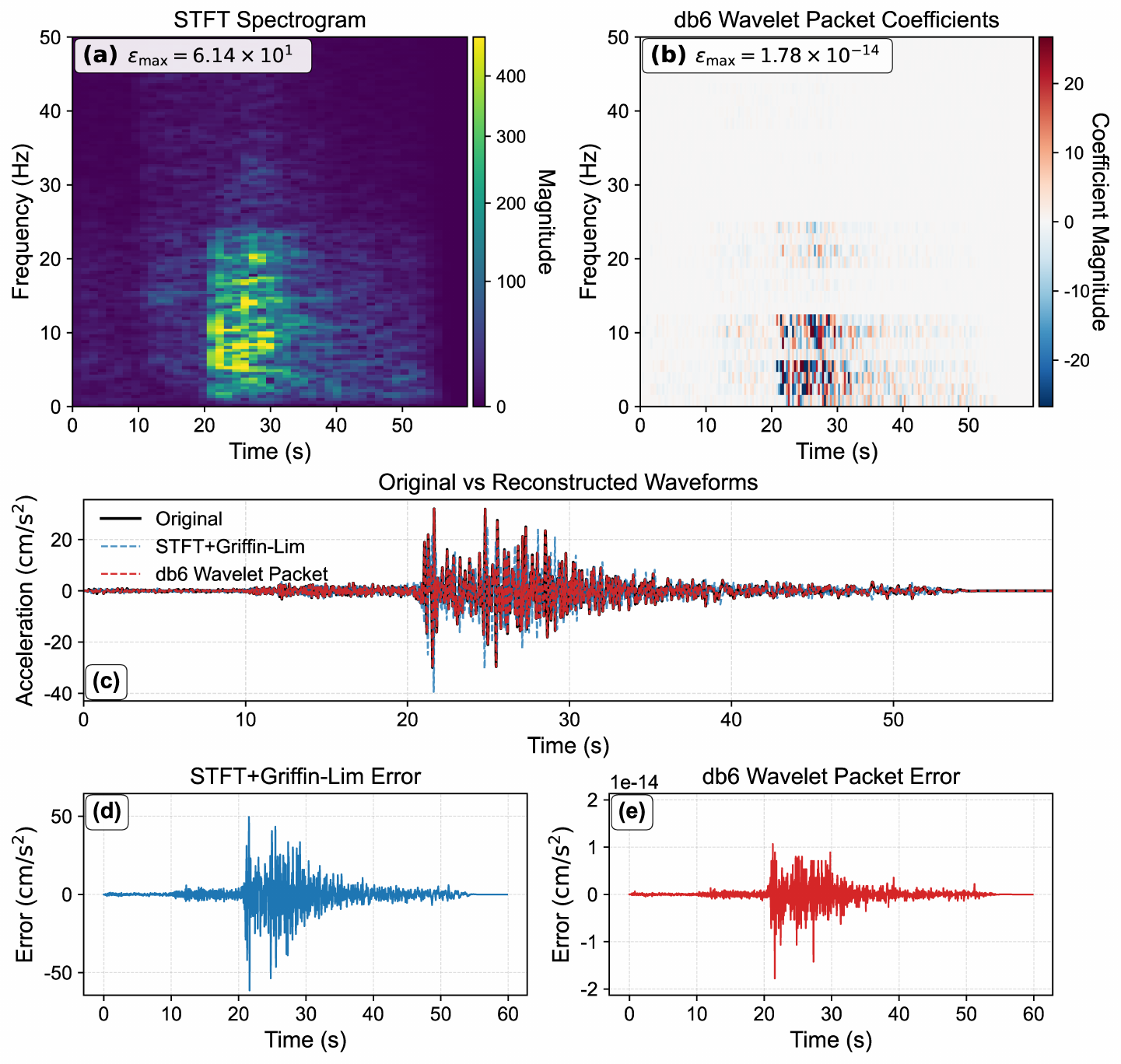}
 \caption{Comparison between STFT+Griffin-Lim and wavelet packet reconstruction. (a) STFT spectrogram with maximum reconstruction error $\varepsilon_{\mathrm{max}} = 6.14 \times 10^{1}$; (b) Wavelet packet coefficients with $\varepsilon_{\mathrm{max}} = 1.78 \times 10^{-14}$; (c) original waveform (black) compared with STFT+Griffin-Lim reconstruction (blue dashed) and wavelet packet reconstruction (red dashed); (d) STFT+Griffin-Lim reconstruction error; (e) Wavelet packet reconstruction error.}
 \label{fig:stft_vs_db6}
\end{figure}

\begin{figure*}
 \centering
 \includegraphics[width=0.95\textwidth]{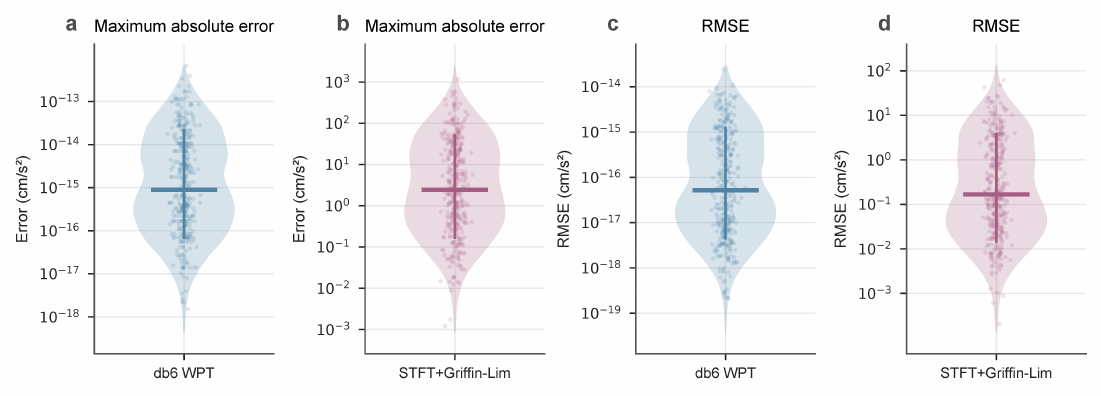}
 \caption{Reconstruction error distributions of the Db6 wavelet packet and STFT+Griffin-Lim representations across the test set. (a, b) Maximum absolute reconstruction error. (c, d) Waveform RMSE.}
 \label{fig:dataset_recon_stats}
\end{figure*}

Compared with the conventional STFT magnitude representation, the wavelet packet transform preserves phase and enables reconstruction through a single inverse transform, avoiding the iterative phase recovery required by Griffin-Lim. Figure~\ref{fig:stft_vs_db6} shows a representative comparison. Across all test records, the wavelet packet round-trip error remained at the numerical round-off level, with a maximum absolute error below $7 \times 10^{-13}$ cm/s$^2$ and a median waveform RMSE of $5.2 \times 10^{-17}$ cm/s$^2$. In contrast, the per-record maximum absolute reconstruction error for STFT+Griffin-Lim averaged 35.88 cm/s$^2$ across the test set, and the mean waveform RMSE was 2.53 cm/s$^2$ (Fig.~\ref{fig:dataset_recon_stats}).

\subsection{Diffusion Model Theory}
\label{subsec:diffusion}

Diffusion models learn data distributions through progressive noising and denoising. This study adopts the Elucidating Diffusion Model (EDM) framework of Karras et al. \citep{Karras_2022_ElucidatingDesignSpace} and extends it by representing ground motions in wavelet packet space and incorporating heterogeneous conditioning variables. EDM defines a design space for diffusion models by decoupling the noise schedule, network preconditioning, loss weighting, and sampling algorithm.

Let the data distribution be denoted as $p_\text{data}(\mathbf{x})$ with variance $\sigma_\text{data}^2$. For conditional generation tasks, the data distribution is represented as a mixture distribution of conditional parameters:
\begin{equation}
p_\text{data}(\mathbf{x})=\mathbb{E}_{p_\text{cond}(\mathbf{c})}[p_\text{data}(\mathbf{x}|\mathbf{c})],
\label{eq:data_dist}
\end{equation}
where $\mathbf{c}$ denotes the conditional parameters. Adding Gaussian noise with variance $\sigma^2$ yields a family of noisy distributions $p(\mathbf{x};\sigma)$. When the noise standard deviation is sufficiently large ($\sigma_\text{max} \gg \sigma_\text{data}$), $p(\mathbf{x};\sigma_\text{max})$ approximates a pure noise distribution.

For the forward noising process, adding Gaussian noise with standard deviation $\sigma$ to clean data $\mathbf{x}_0$ yields the noisy sample:
\begin{equation}
\mathbf{x}_t = \mathbf{x}_0 + \sigma \epsilon, \quad \epsilon \sim \mathcal{N}(0, \mathbf{I}),
\label{eq:forward_sample}
\end{equation}
where $t$ denotes a general index for the noisy state, $\sigma$ represents the noise level, and the transition kernel is $p(\mathbf{x}_t|\mathbf{x}_0) = \mathcal{N}(\mathbf{x}_t | \mathbf{x}_0, \sigma^2 \mathbf{I})$.

The core concept of diffusion models is to start from pure Gaussian noise sampled at the maximum noise level $\sigma_\text{max}$ and progressively denoise to generate clean samples conforming to the target distribution $p_\text{data}(\mathbf{x}|\mathbf{c})$. We discretize the denoising process into $T$ steps, starting from $\mathbf{x}_T \sim \mathcal{N}(0,\sigma_\text{max}^2\mathbf{I})$ and iteratively updating to $\mathbf{x}_0$. The noise level $\sigma$ is sampled from a log-normal distribution ($\ln \sigma \sim \mathcal{N}(P_\text{mean}, P_\text{std}^2)$) to balance learning across noise scales. The range $\sigma \in [0.002, 80]$, the sampling parameters ($P_\text{mean} = -1.2$, $P_\text{std} = 1.2$), $\sigma_\text{data} = 0.5$, and the weighting function $\lambda(\sigma)$ were set to the reference values reported by Karras et al. \citep{Karras_2022_ElucidatingDesignSpace}.

During training, the diffusion model employs the denoising score matching objective function. This approach circumvents computation of normalization constants by directly learning the score function $\nabla_\mathbf{x}\log p(\mathbf{x};\sigma)$ from noisy data. The denoising function $D(\mathbf{x};\sigma)$ minimizes the $L_2$ denoising error:
\begin{equation}
\mathcal{L} = \mathbb{E}_{\sigma,\boldsymbol{y},\boldsymbol{n}}\left[\lambda(\sigma)\left\|D(\boldsymbol{y}+\boldsymbol{n};\sigma)-\boldsymbol{y}\right\|_2^2\right],
\label{eq:loss}
\end{equation}
where $\sigma \sim p_{\text{train}}$, $\boldsymbol{y}\sim p_\text{data}$, $\boldsymbol{n}\sim\mathcal{N}(\mathbf{0},\sigma^2\mathbf{I})$. The weighting function $\lambda(\sigma) = (\sigma^2 + \sigma_{\text{data}}^2)/(\sigma \cdot \sigma_{\text{data}})^2$ (with $\sigma_{\text{data}} = 0.5$ in this study) ensures uniform loss weighting across different noise levels, avoiding training bias toward specific noise scales.
Thus the score matching relation holds:
\begin{equation}
\nabla_\mathbf{x}\log p(\mathbf{x};\sigma)=\frac{D(\mathbf{x};\sigma)-\mathbf{x}}{\sigma^2}.
\end{equation}

The key observation of diffusion models is that $D(\mathbf{x};\sigma)$ can be trained using a neural network $D_{\theta}(\mathbf{x};\sigma)$. To maintain network inputs and outputs within stable numerical ranges, EDM introduces skip connections and scaling preconditioning mechanisms. The denoising network $D_\theta$ is constructed as:
\begin{equation}
D_\theta(\mathbf{x}, \sigma) = c_{\text{skip}}(\sigma) \mathbf{x} + c_{\text{out}}(\sigma) F_\theta(c_{\text{in}}(\sigma) \mathbf{x}, c_{\text{noise}}(\sigma)),
\label{eq:preconditioning}
\end{equation}
where $F_\theta$ denotes the U-Net backbone network. $c_{\text{skip}}(\sigma)$ modulates the skip connection, $c_{\text{in}}(\sigma)$ and $c_{\text{out}}(\sigma)$ scale the input and output magnitudes, and $c_{\text{noise}}(\sigma)$ maps noise level $\sigma$ into a conditioning input for $F_\theta$.

The scaling functions are derived through the following principles: (1) network inputs possess unit variance; (2) training targets possess unit variance; (3) minimization of network error amplification coefficients; (4) uniform distribution of loss weights across different noise levels. The derivation yields:
\begin{equation}
\begin{cases}
c_{\text{skip}}(\sigma)&=\sigma_\text{data}^2/(\sigma^2+\sigma_\text{data}^2) \\
c_{\text{out}}(\sigma)&=\sigma\cdot\sigma_\text{data}/\sqrt{\sigma_\text{data}^2+\sigma^2} \\
c_{\text{in}}(\sigma)&=1/\sqrt{\sigma_\text{data}^2+\sigma^2} \\
c_{\text{noise}}(\sigma)&=\frac{1}{4}\ln(\sigma)
\end{cases}
\label{eq:scaling}
\end{equation}

We can equivalently express the loss (Eq. \ref{eq:loss}) with respect to the raw network output $F_{\theta}$ in Eq. \ref{eq:preconditioning}:
\begin{equation}
\mathcal{L} = \mathbb{E}_{\sigma,\boldsymbol{y},\boldsymbol{n}}
[\underbrace{\lambda(\sigma)c_{\mathrm{out}}(\sigma)^{2}}_{\text{effective weight}}\Vert\underbrace{F_{\theta}\left(c_{\mathrm{in}}(\sigma)\cdot(\boldsymbol{y+n});c_{\mathrm{noise}}(\sigma)\right)}_{\text{network output}}-\underbrace{\frac{1}{c_{\mathrm{out}}(\sigma)}\left(\boldsymbol{y}-c_{\mathrm{skip}}(\sigma)\cdot(\boldsymbol{y+n})\right)}_{\text{effective training target}}\Vert_{2}^{2}
].
\label{eq:loss_re}
\end{equation}

During inference, starting from Gaussian noise $\mathbf{x}_T \sim \mathcal{N}(0, \sigma_\text{max}^2 \mathbf{I})$, a second-order Heun method is used to solve the ODE. Compared to the first-order Euler method, this significantly reduces discretization error for the same number of steps. This study adopts $N=25$ noise levels (corresponding to 50 network evaluations), substantially enhancing efficiency while maintaining generation quality compared to traditional DDPMs (1000 steps). This deterministic sampling yields unique solutions for a given condition and initial noise, facilitating subsequent analysis. Conditional sample variability can be characterized by varying the initial noise while holding the conditioning inputs fixed.

\subsection{Network Architecture}

The core of the diffusion model is the parameterized denoising network for predicting clean data. The denoising network operates directly in wavelet packet coefficient space. Its input and output are $128\times128$ matrices, with rows representing frequency subbands and columns representing time indices. The convolutional kernels therefore learn local time-frequency energy patterns, while the conditioning mechanism described in Section~\ref{subsec:condition} guides the time-frequency distribution of energy in the generated coefficients.

This work adopts a U-Net-based architecture, comprising an encoder (downsampling path), a bottleneck layer, and a decoder (upsampling path), with skip connections that enable multi-scale feature fusion. The network uses 64 base channels with multipliers [1, 2, 4, 4], incorporating 2 residual blocks per resolution level. Self-attention is applied only at the eightfold downsampled resolution, where the $16\times16$ feature maps contain 256 tokens. At this resolution, the model captures global time-frequency interactions at tractable computational cost, following the coarse-resolution attention design of the DDPM U-Net \citep{Ho_2020_DenoisingDiffusionProbabilistic}. This configuration balances model expressiveness with computational efficiency.

The network body consists of stacked residual blocks and attention blocks. Residual blocks incorporate normalization, activation functions, and convolution operations to achieve nonlinear feature transformations. Attention blocks integrate two attention mechanisms: self-attention and conditional cross-attention. Self-attention allows time-frequency positions within features to interact, learning long-range time-frequency dependencies. Cross-attention allows waveform features in the denoising process to interact with encoded condition tokens, providing conditional control over the intensity measures and spectral characteristics of the generated waveforms.

The denoising network is conditioned on the current noise level. The noise level is first transformed as $c_{\text{noise}}(\sigma) = \frac{1}{4} \ln \sigma$ (Eq.~\ref{eq:scaling}) and then mapped to a high-dimensional embedding using sinusoidal positional encoding:
\begin{equation}
\text{PE}(c_{\text{noise}}) = [\sin(\omega_1 c_{\text{noise}}), \cos(\omega_1 c_{\text{noise}}), \dots, \sin(\omega_{d/2} c_{\text{noise}}), \cos(\omega_{d/2} c_{\text{noise}})],
\label{eq:positional_encoding}
\end{equation}
where $\omega_i = 1/10000^{2i/d}$ for $i \in \{0, 1,..., d/2-1\}$ denotes a predefined frequency, and $d$ is the embedding dimension. This encoding provides a smooth representation across noise scales. Within the U-Net residual blocks, an MLP projects the noise embedding to the number of feature channels, and the projected embedding is added elementwise to the feature map to condition the denoising operation on $\sigma$.

\subsection{Cross-Attention Conditional Injection}
\label{subsec:condition}

To control the intensity measures and time-frequency characteristics of the generated seismic waveforms, the conditioning variables are encoded in a form that can be injected into the diffusion process. MLPs project scalar conditions, such as Arias intensity, into individual tokens. Vector conditions, such as the response spectrum and the Husid curve, are partitioned into patches and mapped to token sequences using MLPs. Positional encodings are added to the tokens of each vector condition to represent their order. For the response spectrum, token positions progress from short to long periods. For the Husid curve, they progress from early to late stages of energy accumulation.

To model dependencies among the conditioning variables, we use a three-layer Transformer encoder. Each layer contains a multi-head self-attention module with four attention heads and a position-wise feed-forward network. The layers allow the condition tokens to interact and encode their global dependencies.

We introduce a cross-attention mechanism into the attention blocks of the U-Net encoder, bottleneck layer, and decoder. Cross-attention enables interaction between waveform features and conditional token sequences during the denoising process, achieving fine-grained conditional control over generation. We first flatten the input wavelet packet time-frequency representation into a sequence. For a time-frequency representation containing temporal and frequency information $\mathbf{x} \in \mathbb{R}^{C \times H \times W}$, this operation converts the two-dimensional feature map into a linear sequence $\mathbf{x}_{\text{seq}} \in \mathbb{R}^{(H \cdot W) \times C}$. Then, this waveform feature sequence is projected to obtain the query matrix:
\begin{equation}
\mathbf{Q} = \mathbf{x}_{\text{seq}} \mathbf{W}_Q \in \mathbb{R}^{(H \cdot W) \times d_k}.
\label{eq:query}
\end{equation}

The condition-token sequence $\mathbf{C}_{\text{mixed}}$ output by the conditional encoder is linearly projected to obtain the key ($\mathbf{K}$) and value ($\mathbf{V}$) matrices:
\begin{equation}
\mathbf{K} = \mathbf{C}_{\text{mixed}} \mathbf{W}_K \in \mathbb{R}^{N_c \times d_k}, \quad
\mathbf{V} = \mathbf{C}_{\text{mixed}} \mathbf{W}_V \in \mathbb{R}^{N_c \times d_v},
\label{eq:key_value}
\end{equation}
where $N_c$ denotes the total number of conditional tokens, and $\mathbf{W}_Q, \mathbf{W}_K, \mathbf{W}_V$ are learnable projection matrices.

The attention weights are computed from the scaled dot product between the waveform-feature queries and condition-token keys:
\begin{equation}
\text{Attn} = \text{softmax}\left(\frac{\mathbf{Q} \mathbf{K}^T}{\sqrt{d_k}}\right) \in \mathbb{R}^{(H \cdot W) \times N_c}.
\label{eq:attention}
\end{equation}

The attention weights are used to weight the value vectors, yielding $\text{Attn} \cdot \mathbf{V} \in \mathbb{R}^{(H \cdot W) \times d_v}$. The attention output is then linearly projected to match the channel dimension $C$, reshaped to $\mathbb{R}^{C \times H \times W}$, and added to the original features through a residual connection.

We use multi-head attention with four heads to model interactions between condition tokens and waveform features. The channel dimension $C$ is evenly divided into four subspaces, and each head independently performs scaled dot-product attention to produce an output of dimension $d_v = C/4$. The outputs of the four heads are concatenated and linearly projected back to the original channel dimension, allowing the model to capture multiple conditional dependencies in parallel.

\section{Database Construction and Preprocessing}

The training and testing datasets were constructed from the NGA-West2 database \citep{Ancheta_2014_NGAWest2Database} released by the Pacific Earthquake Engineering Research Center (PEER). Following PEER's quality control, we performed additional data-integrity and validity checks, yielding 19,760 seismic acceleration records comprising the first horizontal component of each record. To evaluate the model's conditional generation capability for seismic events not involved in training, it should be avoided that related records of the same event enter both the training set and the test set simultaneously. In this paper, the dataset is divided according to seismic events. Specifically, all 422 events are split by event count in a 9:1 ratio into 380 training events and 42 test events, containing 17,784 training records and 1,976 test records, respectively.

The data preprocessing pipeline comprised (1) resampling to 100 Hz, for which records with lower sampling rates were upsampled by linear interpolation and those with higher sampling rates were low-pass filtered and downsampled, (2) standardizing each record to 16,384 samples (163.84 s) by truncation or zero-padding to match the $128 \times 128$ wavelet packet representation described in Section~\ref{subsec:wavelet}, and (3) computing and normalizing the response spectra and selected intensity measures.

Under this preprocessing, the wavelet packet representation nominally spans 0--50 Hz, but the frequency range common to the full dataset is limited by the original recordings. The lower bound is determined by the record-specific high-pass corner frequencies applied during PEER processing \citep{Ancheta_2014_NGAWest2Database}, whereas the upper bound is determined by the original sampling rates. The original Nyquist frequency is at least 25 Hz for all records and at least 50 Hz for 92.1\% of the records. Accordingly, the common effective frequency band is approximately 0.1--25 Hz, which covers the principal range of engineering interest, while most records retain information up to 40--50 Hz.

The conditioning variables describe spectral shape and temporal-energy characteristics. They comprise scalar intensity measures, including Arias intensity ($I_A$), 5\% energy arrival time ($T_5$), significant duration ($D_{5-95}$), and temporal centroid ($E_{th}$), as well as vector conditions comprising the response spectrum and the Husid curve.
Arias intensity reflects the total seismic energy, defined as:
\begin{equation}
I_A = \frac{\pi}{2g} \int_0^{T_d} a^2(t) \, dt,
\label{eq:arias}
\end{equation}
where $a(t)$ denotes the acceleration time history, $T_d$ is the total duration, and $g$ is gravitational acceleration. The time $T_5$ is defined as the instant at which the normalized cumulative Arias intensity reaches 5\%. Significant duration $D_{5-95}$ is the interval between $T_5$ and the corresponding 95\% time $T_{95}$, i.e., $D_{5-95} = T_{95} - T_5$. This interval characterizes the principal energy-release phase of the ground motion.

The temporal centroid ($E_{th}$) describes the temporal center of seismic energy release \citep{Yamamoto_2013_StochasticModelEarthquake}, defined as:
\begin{equation}
E_{th} = \frac{\int_0^{T_d} t \cdot a^2(t) \, dt}{\int_0^{T_d} a^2(t) \, dt},
\label{eq:E_th}
\end{equation}
where the numerator is the time integral weighted by the square of acceleration, and the denominator represents the total energy.

The normalized Husid curve represents the temporal evolution of cumulative Arias intensity \citep{Huang_2017_EnergycompatibleSpectrumcompatibleECSC}. Because Arias intensity is included separately as a scalar condition, the Husid curve describes the pattern of energy accumulation:
\begin{equation}
H(t) = \frac{\int_0^t a^2(\tau) \, d\tau}{\int_0^{T_d} a^2(\tau) \, d\tau} \in [0,1].
\label{eq:husid}
\end{equation}

The Husid curve is monotonic and smooth. It was therefore downsampled to 256 points uniformly distributed over $[0,T_d]$ to balance representation efficiency and feature completeness.

To reduce scale differences among the conditioning variables, all variables were normalized. Because Arias intensity spans several orders of magnitude, a logarithmic transformation was combined with min-max scaling. The transformation $u = \log(1 + x)$ was first applied to compress the dynamic range, after which the values were mapped to $[-1, 1]$:
\begin{equation}
x_{\text{norm}} = 2 \cdot \frac{u - u_{\text{min}}}{u_{\text{max}} - u_{\text{min}}} - 1,
\label{eq:log_minmax}
\end{equation}
where $u_{\text{min}}$ and $u_{\text{max}}$ denote the minimum and maximum values after logarithmic transformation, respectively. This approach compressed the dynamic range while preserving monotonicity. The same min-max scaling formula (Eq.~\ref{eq:log_minmax}) was applied to $T_5$, $D_{5-95}$, and $E_{th}$ without logarithmic transformation.

The response spectrum $S_a(T)$ of each record was normalized by PGA to preserve its spectral shape. The Husid curve is dimensionless and bounded between 0 and 1 and was therefore used directly as a conditioning variable.

\section{Experimental Results and Analysis}

To evaluate the effect of the conditioning information on generation quality, we compared two conditioning configurations. Case 1 used only the response spectrum normalized by PGA. Case 2 used the response spectrum ($S_a$), Arias intensity ($I_A$), 5\% energy arrival time ($T_5$), significant duration ($D_{5-95}$), temporal centroid ($E_{th}$), and Husid curve ($H(t)$).

The models were trained using the Adam optimizer with a learning rate of $1 \times 10^{-4}$ and a batch size of 110. To stabilize the generation quality, an exponential moving average (EMA) was applied to the model parameters with a decay rate of 0.999. Training was conducted in a distributed manner across three NVIDIA A100 GPUs employing the Distributed Data Parallel (DDP) strategy and mixed precision (BF16) to enhance computational efficiency. Each model was trained for 100 epochs, requiring approximately 2.5 hours to complete. For inference, we utilized the second-order Heun sampler with 25 steps (equating to 50 network evaluations) to achieve a balance between sampling efficiency and output quality. The inference was performed on a single NVIDIA A100 GPU with a batch size of 256, processing the entire test set in approximately 3 minutes.

\subsection{Control Accuracy of Scalar Intensity Measures}

\begin{figure*}
 \centering
 \includegraphics[width=0.9\textwidth]{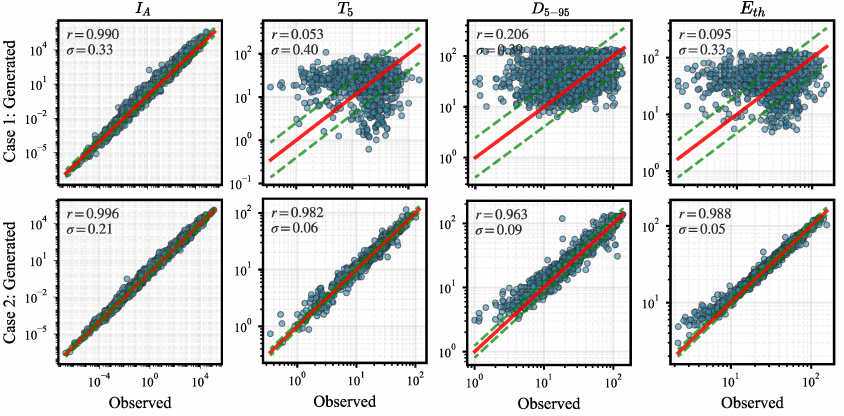}
 \caption{Comparison of scalar intensity-measure control accuracy between Case 1 and Case 2.}
 \label{fig:ims_compare}
\end{figure*}

We performed single-pass inference, with one sample per test condition, to evaluate the control accuracy of scalar intensity measures. Figure~\ref{fig:ims_compare} compares the intensity measures obtained with the Case 1 and Case 2 models. Solid red lines indicate exact agreement, dashed green lines indicate $\pm$ one residual standard deviation, and the annotations report $r$ (Pearson correlation) and $\sigma$ (residual standard deviation).

For Arias intensity ($I_A$), Case 1 achieved a correlation of $r=0.990$ and a residual standard deviation of $\sigma=0.33$, although $I_A$ was not explicitly included as a condition. This result suggests that the model may infer overall energy characteristics from PGA and the response spectrum. In Case 2, which explicitly included $I_A$, the correlation increased to $r=0.996$ and the residual standard deviation decreased to $\sigma=0.21$, indicating improved control of total energy.

Performance differences between the two cases were most pronounced for the temporal-energy measures. For the 5\% energy arrival time ($T_5$), Case 1 yielded $r=0.053$ with a residual standard deviation of $\sigma=0.40$, indicating that the model did not accurately reproduce the energy-onset time without explicit temporal-energy conditions. Under the complete Case 2 conditioning set, the correlation increased to $r=0.982$ and the residual standard deviation decreased to $\sigma=0.06$. The results for $D_{5-95}$ and $E_{th}$ followed similar trends. These results indicate that Case 2 provides more accurate control of temporal-energy characteristics than Case 1 across the entire NGA-West2 test set.

\begin{figure*}
 \centering
 \includegraphics[width=0.9\textwidth]{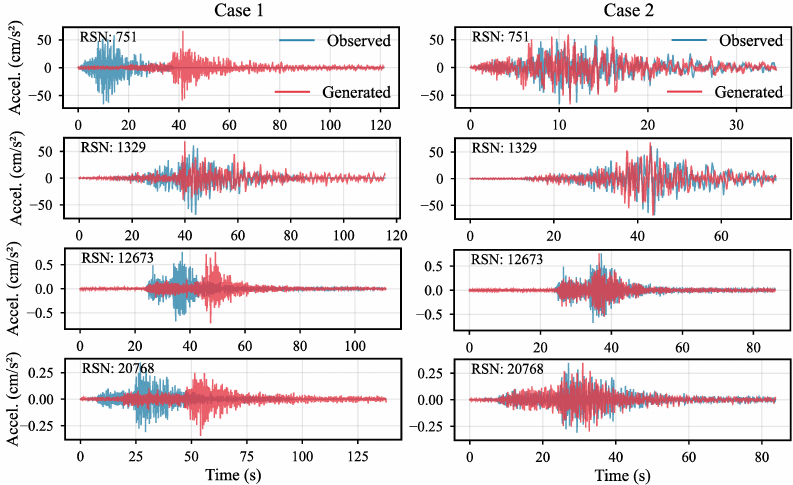}
 \caption{Comparison of generated and observed waveforms for eight randomly selected test records from Case 1 and Case 2.}
 \label{fig:case1and2_waveform}
\end{figure*}

For further validation, we randomly selected eight test records for waveform visualization. Figure~\ref{fig:case1and2_waveform} compares the generated and observed waveforms for both cases up to the time at which the cumulative Arias intensity reaches 99.5\%. RSN denotes the unique record sequence number assigned to each record in the NGA-West2 database. In Case 1, the generated amplitude envelopes generally followed those of the observed waveforms, with the principal energy concentrated in the central portions and decaying toward both ends. However, the generated energy-onset times deviated from the observations, consistent with the low control accuracy for $T_5$ in Figure~\ref{fig:ims_compare}.

Under the complete Case 2 conditioning set, the generated waveforms showed closer agreement with the observations in both amplitude envelope and temporal-energy distribution. The energy-onset times, principal energy-containing intervals, and decay phases more closely matched the observations. These comparisons indicate that Case 2 reduced the timing offsets and dispersion in energy accumulation observed for Case 1.

Table~\ref{tab:accuracy_samples} presents quantitative results for the eight randomly selected test records under both conditioning configurations. For response-spectrum matching, both cases demonstrate excellent performance with $R^2$ values generally exceeding 0.95; however, for Husid-curve matching, Case 2 exhibits significant advantages over Case 1, with all samples achieving $R^2$ values exceeding 0.98.

\captionsetup{labelformat=empty}
\begin{table*}[htbp]
 \centering
 \caption{\centering \textbf{Table~\thetable}: Response-spectrum and Husid-curve matching accuracy for eight randomly selected test records}
 \label{tab:accuracy_samples}
 \begin{tabularx}{\textwidth}{c*{8}{>{\centering\arraybackslash}X}}
 \toprule
 \multirow{2}{*}{RSN} & \multicolumn{2}{c}{Case 1 $S_a$} & \multicolumn{2}{c}{Case 2 $S_a$} & \multicolumn{2}{c}{Case 1 $H(t)$} & \multicolumn{2}{c}{Case 2 $H(t)$} \\
 \cmidrule(lr){2-3} \cmidrule(lr){4-5} \cmidrule(lr){6-7} \cmidrule(lr){8-9}
 & $R^2$ & RMSE & $R^2$ & RMSE & $R^2$ & RMSE & $R^2$ & RMSE \\
 \midrule
 635 & 0.976 & 0.284 & 0.979 & 0.266 & -0.309 & 0.364 & 0.984 & 0.022 \\
 2308 & 0.960 & 0.236 & 0.972 & 0.199 & 0.295 & 0.296 & 0.990 & 0.035 \\
 3572 & 0.950 & 0.215 & 0.958 & 0.197 & 0.084 & 0.265 & 0.997 & 0.015 \\
 8403 & 0.990 & 0.197 & 0.989 & 0.212 & 0.580 & 0.240 & 0.998 & 0.015 \\
 14702 & 0.983 & 0.346 & 0.988 & 0.287 & 0.909 & 0.123 & 0.997 & 0.021 \\
 15278 & 0.996 & 0.150 & 0.996 & 0.150 & 0.914 & 0.121 & 0.996 & 0.025 \\
 18231 & 0.993 & 0.134 & 0.993 & 0.134 & 0.933 & 0.109 & 0.998 & 0.021 \\
 20334 & 0.991 & 0.159 & 0.988 & 0.185 & 0.882 & 0.131 & 0.994 & 0.030 \\
 \bottomrule
 \end{tabularx}
\end{table*}
\captionsetup{labelformat=default}

To determine whether the advantage of the wavelet packet representation extends beyond reconstruction accuracy to generative performance, we additionally trained an STFT magnitude baseline for Case 2 while keeping all other settings unchanged. The generated magnitude spectrograms were converted to waveforms using 128 Griffin-Lim iterations. Both models achieved comparable control of Arias intensity, but the wavelet packet model yielded higher $r$ and lower $\sigma$ for $T_5$, $D_{5-95}$, and $E_{th}$ (Table~\ref{tab:stft_baseline}). The end-to-end generation time was approximately 0.0837 s per record for WPT and 0.2708 s per record for STFT+Griffin-Lim. STFT+Griffin-Lim therefore had an end-to-end generation time approximately 3.2 times that of WPT (Table~\ref{tab:stft_timing}).

\captionsetup{labelformat=empty}
\begin{table*}[htbp]
 \centering
 \caption{\centering \textbf{Table~\thetable}: Comparison of scalar intensity-measure control performance between WPT and STFT+Griffin-Lim}
 \label{tab:stft_baseline}
 \begin{tabular}{ccccc}
 \toprule
 \multirow{2}{*}{Parameter} & \multicolumn{2}{c}{WPT} & \multicolumn{2}{c}{STFT+Griffin-Lim} \\
 \cmidrule(lr){2-3} \cmidrule(lr){4-5}
 & $r$ & $\sigma$ & $r$ & $\sigma$ \\
 \midrule
 $I_A$ & 0.996 & 0.21 & 0.994 & 0.25 \\
 $T_5$ & 0.982 & 0.06 & 0.809 & 0.21 \\
 $D_{5-95}$ & 0.963 & 0.09 & 0.752 & 0.25 \\
 $E_{th}$ & 0.988 & 0.05 & 0.822 & 0.18 \\
 \bottomrule
 \end{tabular}
\end{table*}

\begin{table*}[htbp]
 \centering
 \caption{\centering \textbf{Table~\thetable}: Comparison of per-record computation time between WPT and STFT+Griffin-Lim}
 \label{tab:stft_timing}
 \begin{tabular}{lcc}
 \toprule
 Timing component & WPT & STFT+Griffin-Lim \\
 \midrule
 Network sampling per record (s) & 0.0816 & 0.0782 \\
 Reconstruction per record (s) & 0.0021 & 0.1926 \\
 End-to-end per record (s) & 0.0837 & 0.2708 \\
 \bottomrule
 \end{tabular}
\end{table*}
\captionsetup{labelformat=default}

\begin{figure*}
 \centering
 \includegraphics[width=0.9\textwidth]{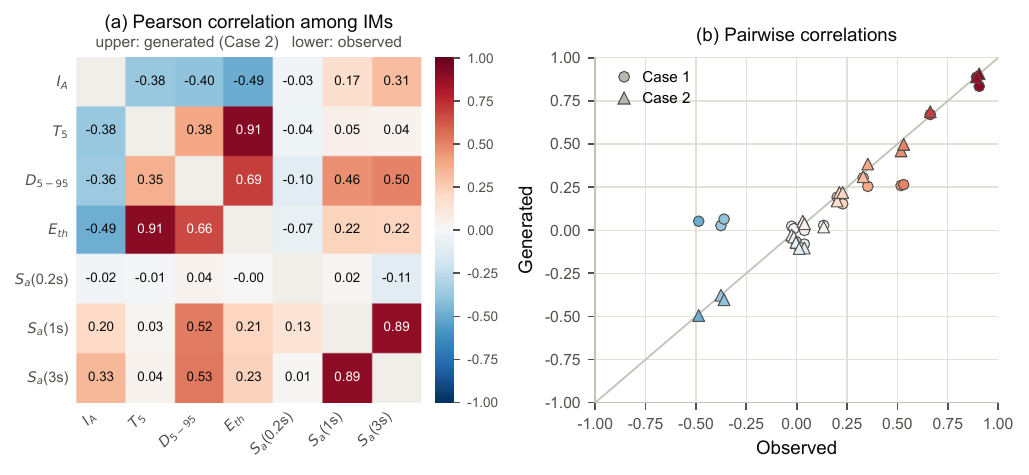}
 \caption{Pairwise Pearson correlations among seven log-transformed intensity measures of the observed and generated records. (a) Combined correlation matrix for the observed records (lower triangle) and Case 2 (upper triangle). (b) Generated versus observed pairwise correlations for Cases 1 and 2. Marker color encodes the corresponding Case 2 correlation according to the colormap in (a).}
 \label{fig:im_correlation}
\end{figure*}

To examine whether the generated motions preserve the dependence among spectral shape, energy content, and duration, Fig.~\ref{fig:im_correlation} compares pairwise Pearson correlations among seven log-transformed intensity measures in the observed and generated test sets. These measures are $I_A$, $T_5$, $D_{5-95}$, $E_{th}$, and $S_a(T)/\mathrm{PGA}$ at $T=0.2$, 1, and 3 s. In Fig.~\ref{fig:im_correlation}(a), the lower triangle reports the observed correlations and the upper triangle reports those from Case 2.

Case 2 reproduced the empirical correlation structure, with a mean absolute deviation of 0.036 across the 21 pairs and a maximum absolute deviation of 0.141. This agreement indicates that the empirical dependence among the measures was broadly preserved. Case 1 retained most relationships but not the negative associations between $I_A$ and the duration-related measures. Correlations of approximately $-0.4$ in the observed set were close to zero in the Case 1 generated set. Explicit temporal-energy conditioning in Case 2 largely recovered this energy-duration coupling.

\subsection{Analysis of Conditional Sample Variability}

Diffusion models can generate diverse samples under fixed conditions. Here, conditional sample variability denotes the sample-to-sample dispersion of the learned conditional distribution $p(\mathbf{x}|\mathbf{c})$ when the conditioning inputs are fixed and only the initial noise is varied. In this section, 150 condition sets were randomly selected from the test dataset, and 100 distinct samples were generated for each condition.

\begin{figure*}
 \centering
 \includegraphics[width=0.85\textwidth]{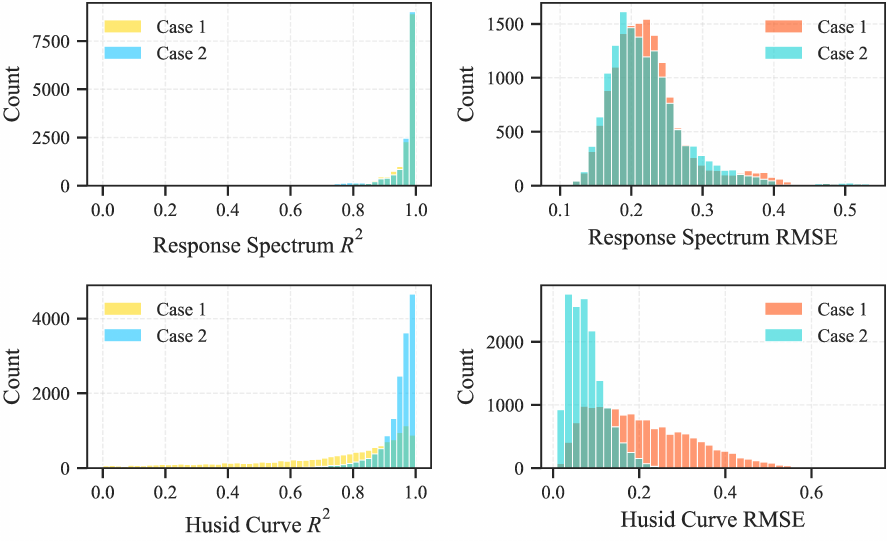}
 \caption{Distribution histograms of accuracy metrics for 15,000 generated samples (150 conditions $\times$ 100 samples). Upper row: response spectrum $R^2$ (left) and RMSE (right) distributions. Lower row: Husid curve $R^2$ (left) and RMSE (right) distributions.}
 \label{fig:accuracy_dist}
\end{figure*}

Figure~\ref{fig:accuracy_dist} presents the distributions of the accuracy metrics for 15,000 generated samples. Accuracy was evaluated using the coefficient of determination ($R^2$) and root-mean-square error (RMSE). For the response spectrum, 91\% of the samples in both cases had $R^2 > 0.9$, and the $R^2$ and RMSE distributions largely overlapped. This result indicates similar response-spectrum matching accuracy under the two conditioning configurations. For the Husid curve, 25\% of the Case 1 samples and 88\% of the Case 2 samples had $R^2 > 0.9$, and Case 2 also showed lower RMSE values. The narrower Case 2 distributions are consistent with its explicit Husid-curve conditioning.

\begin{figure*}
 \centering
 \includegraphics[width=0.85\textwidth]{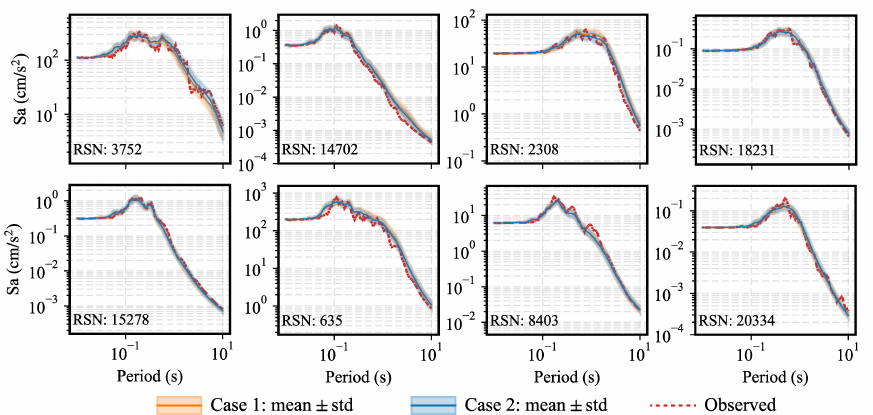}
 \caption{Comparison of variability in response-spectrum matching between Case 1 and Case 2.}
 \label{fig:sa_compare}
\end{figure*}

Figure~\ref{fig:sa_compare} compares Case 1 and Case 2 variability in response-spectrum matching. In both cases, the mean response spectra agreed closely with the observed spectra, and the standard-deviation bands overlapped across most periods. The overlapping bands indicate similar response-spectrum variability for the two cases over most periods.

\begin{figure*}
 \centering
 \includegraphics[width=0.85\textwidth]{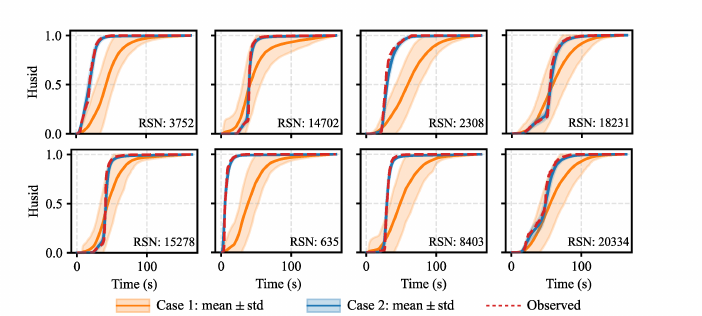}
 \caption{Comparison of Husid curve matching between Case 1 and Case 2. }
 \label{fig:husid_compare}
\end{figure*}

Figure~\ref{fig:husid_compare} further compares conditional sample variability in the generated Husid curves between the two cases. In Case 1, the generated Husid curves preserved the monotonic trend, but the standard-deviation bands widened during the middle stage of energy accumulation, reflecting variability across samples. In Case 2, the mean curves more closely followed the observations and the standard-deviation bands were narrower, consistent with explicit Husid-curve conditioning.

\begin{figure*}
 \centering
 \includegraphics[width=0.85\textwidth]{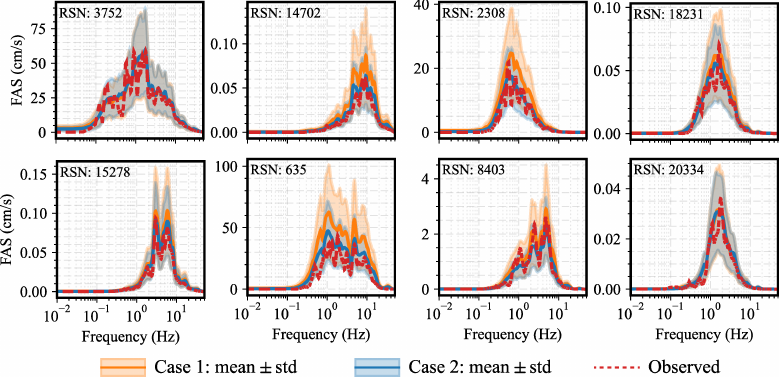}
 \caption{Comparison of Fourier amplitude spectrum matching between Case 1 and Case 2. }
 \label{fig:fas_compare}
\end{figure*}

Figure~\ref{fig:fas_compare} compares Fourier amplitude spectrum (FAS) matching between the two cases. Both cases showed similar agreement with the observed FAS, while Case 2 had slightly narrower standard-deviation bands at intermediate frequencies.

\begin{figure*}
 \centering
 \includegraphics[width=0.85\textwidth]{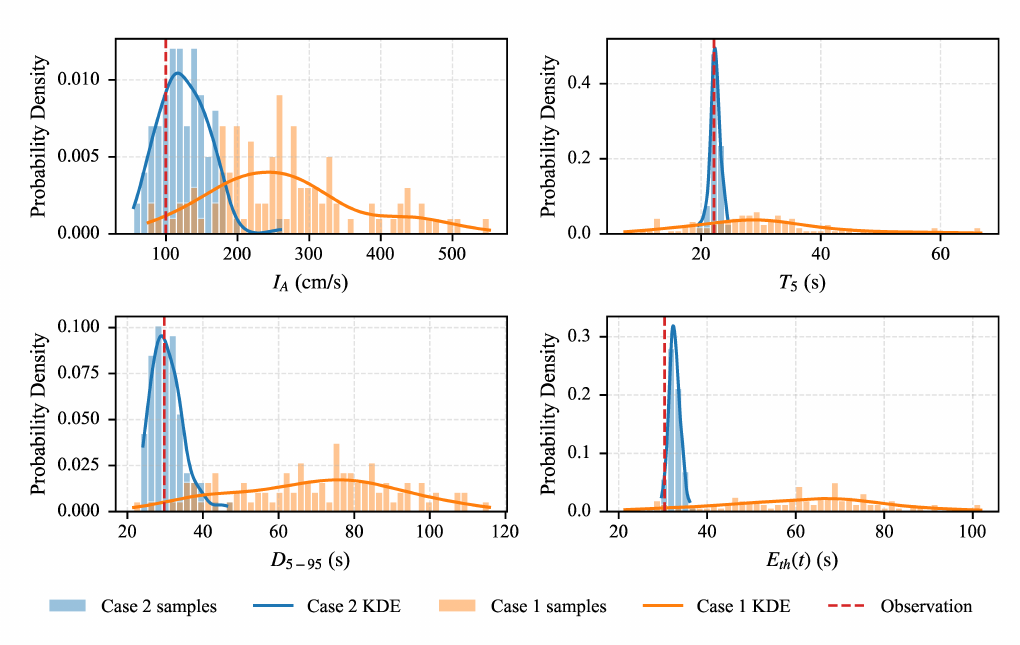}
 \caption{Probability density distributions of intensity measures for RSN 2308 under repeated stochastic sampling (100 samples per case).}
 \label{fig:intensity_pdf_2308}
\end{figure*}

To further quantify the influence of both conditioning configurations on conditional sample variability in the temporal-energy measures, Figure~\ref{fig:intensity_pdf_2308} shows probability density estimates of scalar IMs for the representative record RSN 2308 under stochastic inference with 100 realizations per case. The figure compares Case 1 (orange) and Case 2 (blue) across four measures: Arias intensity ($I_A$), 5\% energy arrival time ($T_5$), significant duration ($D_{5-95}$), and temporal centroid ($E_{th}$). The kernel density estimation (KDE) curves of Case 1 for all four measures exhibit flat profiles spanning wide ranges, indicating substantial dispersion. In contrast, the Case 2 distributions are concentrated near the observed values (red dashed lines) and exhibit markedly less dispersion across all temporal-energy measures. This result indicates that explicit temporal-energy conditioning reduces conditional sample variability.

\section{Discussion}

 The conditioning set deliberately includes partially redundant information. During training and validation, all conditions are extracted from the same record and are therefore internally consistent. The low-dimensional scalars provide the condition encoder with engineering-interpretable summaries of temporal-energy characteristics, whereas the Husid tokens encode the full energy-accumulation profile. The proposed conditional encoding architecture is scalable. The modular Transformer encoder can readily accommodate additional parameters such as peak ground velocity (PGV), cumulative absolute velocity (CAV), or frequency-content metrics through expansion of token sequences. Furthermore, the cross-attention mechanism naturally accommodates condition-token sequences of different lengths, so the conditioning variables can be replaced or extended for a specific application by retraining the model with a new condition set and without modifying the architecture. A trained model, however, operates with a fixed conditioning schema and requires the complete condition set at inference.

For practical engineering applications, condition vectors could be constructed using three strategies. First, temporal-energy conditions could be extracted from a seed record, with only the spectral condition replaced by a target spectrum. This strategy would enable spectrum matching with explicit temporal-energy control. However, the target spectrum and the energy conditions from the seed record should be mutually consistent and lie within the support of the training distribution. The reliability of this strategy requires further validation. Second, scalar conditions could be obtained from published predictive models \citep{Campbell_2019_AriasCAV, Afshari_2016_SignificantDuration} and combined with a parameterized Husid shape. Third, the generalized conditional intensity measure framework \citep{Bradley_2010_GCIM, Bradley_2012_GCIMalgorithm} could be used to sample hazard-consistent condition vectors from a joint conditional distribution of intensity measures, thereby preserving their internal dependence structure.

The 163.84 s analysis window retained, on average, 99.1\% of the Arias intensity of the original records (median, 99.9\%), and 98.3\% of the records reached 95\% of their cumulative energy within this window. The fixed duration balances truncation-related energy loss against information dilution by zero-padding. Padding segments carry no physical information. Excessive padding reduces the effective information density of the coefficient maps and requires the model to allocate part of its learning capacity to uninformative regions. Generative architectures that support variable-length inputs could alleviate this trade-off. One possible solution is a Transformer-based diffusion backbone that handles variable-length coefficient sequences by masking padded positions in both the attention mechanism and the loss function. This possibility warrants further investigation.

The proposed framework is computationally efficient through the variance-exploding diffusion formulation and second-order Heun sampler, requiring only 25 noise levels (50 network evaluations) compared to traditional DDPMs that typically demand 1000 steps. Further acceleration could be achieved with latent diffusion models (LDMs) \citep{Rombach_2021_HighresolutionImageSynthesis}, which perform the denoising process in compressed latent spaces rather than high-dimensional wavelet coefficient spaces. Recent advances in single-step or few-step diffusion models, such as consistency models \citep{Song_2023_ConsistencyModels, Song_2023_ImprovedTechniquesTraining} and progressive distillation techniques \citep{Salimans_2022_ProgressiveDistillationFast}, suggest routes to real-time ground-motion generation. However, the trade-off between inference speed and sample quality warrants careful study, particularly for capturing subtle temporal-energy characteristics that affect structural damage accumulation.

The present validation is limited to the NGA-West2 records used in this study. Performance outside the training distribution remains untested, including extremely long-period pulse-like motions, near-fault directivity effects, and basin-induced surface waves. The fixed time-frequency resolution of the wavelet packet representation may also be less suitable for highly transient motions. Further studies should examine regional transfer, broader motion types, and physics-based constraints for extrapolation beyond the empirical training distribution.

\section{Conclusions}

This study presents a multi-condition diffusion model framework for ground-motion generation that jointly matches the response spectrum and temporal-energy characteristics. By employing wavelet packet representation with perfect reconstruction, Transformer-based conditional encoding, and cross-attention mechanisms, the framework achieves precise control over both spectral and energy-evolution features. Comparative experiments on the NGA-West2 database show that the proposed model learns a probabilistic mapping from conditioning variables to waveforms. The generated motions closely reproduce the record-specific target response spectra and energy-evolution characteristics. Repeated stochastic sampling under fixed conditions was used to quantify conditional sample variability. Future work will focus on constructing hazard-consistent target condition vectors, for example through GCIM-based sampling, and on validating the generated motions in downstream structural response analyses.

\section*{Acknowledgments}

This project is supported by the Intelligent Computing Center of Jianghan University. A preprint version of this manuscript is available on arXiv under the same title (arXiv:2602.07405; \href{https://doi.org/10.48550/arXiv.2602.07405}{https://doi.org/10.48550/arXiv.2602.07405}).

\section*{Conflict of Interest Statement}

The authors declare no conflicts of interest.

\section*{Data Availability Statement}

The NGA-West2 database used in this study is available from the Pacific Earthquake Engineering Research Center (https://peer.berkeley.edu/research/data-sciences/databases, last accessed May 2024). The code will be publicly available on GitHub at https://github.com/CyrusMagic/wavelet-packet-diffusion-gm upon publication.

\bibliography{MyBibList}

\end{document}